# Inverse Opal Optical Tamm State for Sensing Applications

Rina Mudi[a], Alessandro Carpentiero[b], Monica Bollani[c], Mario Barozzi[d], Kapil Debnath[e,f], Andrea Chiappini[b,*], Shivakiran Bhaktha B.N[g,*]

[a]Advanced Technology Development Centre, IIT Kharagpur, West Bengal, 721302, India.

[b]Istituto di Fotonica e Nanotecnologia—CNR, IFN and FBK Photonics Unit, Via alla Cascata 56/c, Povo, 38123 Trento, Italy

[c]Istituto di Fotonica e Nanotecnologie-Consiglio Nazionale Delle Ricerche (IFN-CNR), LNESS Laboratory, 22100 Como, Italy

[d]Centro SD - MTSD , Fondazione Bruno Kessler, Via Sommarive 18, 38123 Trento (Povo), Italy

[e]School of Natural and Computing Sciences, University of Aberdeen, Aberdeen AB243UE, Scotland, United Kingdom

[f]Department of Electronics and Electrical Communication Engineering, IIT Kharagpur, West Bengal, 721302, India.

[g] Department of Physics, IIT Kharagpur, West Bengal, 721302, India.

Corresponding authors: andrea.chiappini@cnr.it, kiranbhaktha@phy.iitkgp.ac.in

**Abstract**

We report the existence of optical Tamm states (OTS) in inverse opal (IO) - based three-dimensional photonic crystal on a flat metal substrate, validated through both numerical simulations and experimental observations. Our fabrication approach for the Tamm inverse opal (Tamm-IO) structure is notably straightforward and does not involve corrosive chemicals. Upon infiltration of non-reactive solvents such as methanol and ethanol into the IO, a noticeable shift of the OTS, consistent with our simulations is observed, and the temporal dynamics of the same have been investigated. The experimentally obtained sensitivity is ~ 110 nm/RIU which is of the same order as the computed value, making the IO OTS to be an attractive sensing tool.

**Keywords:** photonic crystal, inverse opal, surface plasmon resonance, optical Tamm state, Tamm inverse opal, volatile solvents.

1. Introduction:

In the realm of photonics, the manipulation and control of light at the nanoscale have sparked tremendous interest and innovation [1-4], leading to the development of advanced photonic materials [5-7] and devices [8-10] with unprecedented capabilities. Among these, photonic crystals (PhCs) have emerged as a versatile platform offering exquisite control over light propagation, dispersion, and interaction. These periodic nanostructures, exhibiting unique photonic bandgap (PBG) properties akin to their electronic counterparts in semiconductor have found application across a broad spectrum of fields, including sensing [11], imaging [12], and quantum optics [13].
Among the various technique available for manipulating light at the nanoscale, surface electromagnetic waves notably the surface plasmon polaritons (SPPs), have received considerable attention [14-16]. The interest stems from their ability to confine optical field at the interface between metal and dielectric medium. SPPs, in particular have emerged as a highly promising avenue for sensing applications [17], with the surface plasmon resonance (SPR) based refractive index sensors being routinely employed for detecting biomolecules [18-20] and chemicals [21]. The appeal of SPR sensors from their remarkable sensitivity to variation in the surrounding refractive index, has been made possible by the intense confinement of optical fields at the metal-dielectric interface. However, the conventional implementation of SPR sensors typically rely on the Kretschmann-Raether excitation geometry, which involves bulky optical setup based on glass prism to couple light into the SPP mode [22]. Moreover, achieving phase matching between incident beam and the SPP mode often requires relatively high incidence angles. These requirements pose significant challenges for the integration of SPR sensors into miniaturized photonic devices, thus limiting their scalability. In 2007, Kaliteevski et al., [23] proposed an alternative surface electromagnetic wave called the optical Tamm state (OTS) to address the challenges associated with prism coupling. Unlike SPPs, OTS offers the advantage of existing for both TE and TM polarizations and having its dispersion curve lying within the light cone. Consequently, no specialized optical coupling scheme, such as prism coupling or surface grating, is required to excite the plasmon mode. Measurements can be conveniently conducted by simply launching light at normal incidence, making OTS an excellent candidate for integration with nanophotonic devices. But, most of the demonstrations have been limited to one-dimensional PhCs [24-29]. In our earlier work [30], we presented a detailed study on the existence of OTS at the interface of metal-opal structures with different cross-sectional morphologies. We observed the presence of PhC edge-modes and hybrid Tamm-PhC band-edge state in hybrid plasmonic photonic crystal structures. The introduction of IO structure can provide higher porosity for sensing application, which has been explored in this study. Our findings highlight the potential for realizing cost-effective and integrated refractive index sensor based on Tamm IO nanophotonic sensor. An important limitation is that the electric field is confined at the interface of metal and PhC structure. Thus, it is necessary to create opening in the device to allow the analyte access the high field confinement volumes of the structure.

Several attempts have been made to exploit the light confinement in PhCs for sensing applications [25]. Zaky et al., [26] performed computational studies on a refractive index gas sensor based on Tamm state in one-dimensional PhC. Additionally, Zhang et al., [27] proposed a novel concept of refractive index sensing with air-dielectric alternate layered distributed Bragg reflector (DBR) coated with metal. But fabricating such device would require the incorporation of numerous sacrificial layers to attain the desired sub-micron gaps. Moreover, the challenges involved in the fabrication lies in ensuring adequate adhesion between suspended membrane, maintain mechanical stability and addressing potential deformation resulting from residual stress. Juneau-Fecteau et al., [28] reported the fabrication and characterization of a porous Si sensor using Tamm plasmon resonance. But the fabrication of the device consists of a PhC created by periodic electrochemical etching of Si using corrosive chemicals. Burratti et al., [29] reported the temporal response of reflection peak of self-assembled PS opal in presence of various volatile organic compounds (VOCs). But the response time was prolonged, ranging from minutes to hours.

Here, we present the design and development of a Tamm-IO structure aimed at introducing higher porosity in the dielectric region to allow the penetration of the analytes into the device. This is accomplished by substituting the opal structure by $SiO_2$ IO PhC on platinum coated vitreous silica substrate. We observed the excitation of OTS under normal incidence, confirmed by the numerical simulations. For the PhC structures to act as a sensor, it requires the interaction between the analyte and the high field confinement regions, a condition met by IO PhC architecture at the metal-PhC interface. Sensing of VOCs was performed by infiltrating the Tamm-IO with methanol and ethanol and the recovery and response of the OTS were analysed. Notably, the average recovery time for the OTS when ethanol and methanol were infiltrated is found to be $0.4 \pm 0.1$s and $2.0 \pm 0.6$ s, respectively.

## 2. Numerical modelling and analysis

The schematic of Tamm-IO based photonic crystal is shown in fig.1(a). The structure consists of $SiO_2$ IO on platinum (Pt) coated vitreous silica substrate. A total of 12 layers of $SiO_2$ IO with diameter 365 nm and face centered cubic (FCC) stacking were considered along *z*-axis. Bloch boundary conditions were used along the *x* and *y* directions, and perfectly matched layers (PML) were used along the *z*-direction ([111]). The refractive index of SiO$_2$ was considered to be 1.46 [31]. A plane wave propagating along the *z*-axis was incident on the SiO$_2$ IO surface with a wavelength ranging from 400 nm to 900 nm. The simulation time was taken to be 1000 fs with meshing of size 1 nm used along *z*-direction and 10 nm along *x* and *y* directions. In computation, a thin layer of chromium of thickness 5 nm was placed between the 20 nm Pt layer and the vitreous silica substrate, consistent with the fabrication process for promoting adhesion. The normal incidence reflection spectra of the IO and Tamm-IO structures, recorded from the IO side, are shown in Fig. 1(b). The high reflectance observed over the broad band in the spectra of the IO structure corresponds to the band gap of the PhC along the [111] direction. The central wavelength of the PhC bandgap can also be obtained from the modified Bragg's law for first-order diffraction from the [111] planes [32,33]:

$$\lambda_{max} = 2\sqrt{\frac{2}{3}}D\sqrt{n_{void}^2 f + n_{SiO2}^2(1-f) - sin^2\theta} \qquad (1)$$

where, $\lambda_{max}$ is the wavelength at the center of the bandgap, *D* is the diameter of the air-void nanospheres, *f* is the fill fraction of void (*f* = 0.74), *θ* is the angle of incidence, $n_{void}$ and $n_{SiO2}$ are the refractive index of SiO$_2$ (1.46) and void (1.00), respectively. The analytical calculation using Eqn. 1 presents the value of the central wavelength to be 678 nm, which closely matches with the simulated result. In the presence of 20 nm thick Pt layer between the substrate and the SiO$_2$ inverse opal, a dip at ~ 680 nm with an FWHM of ~ 21 nm is observed in the reflection spectrum, within the band gap of the PhC. The electric field distribution at the resonant dip was computed and is shown in Fig. 1(c). The electric field intensity is found to be a maximum at the interface of metal and PhC structure and decays into the depth of the PhC. Additionally, the reflection spectra at various angles of incidence were computed using the 3D-finite difference time domain (FDTD) method, as depicted in the following Fig. 1(d), wherein the blue-shift of the OTS with an increase in the angle of incidence is evident.

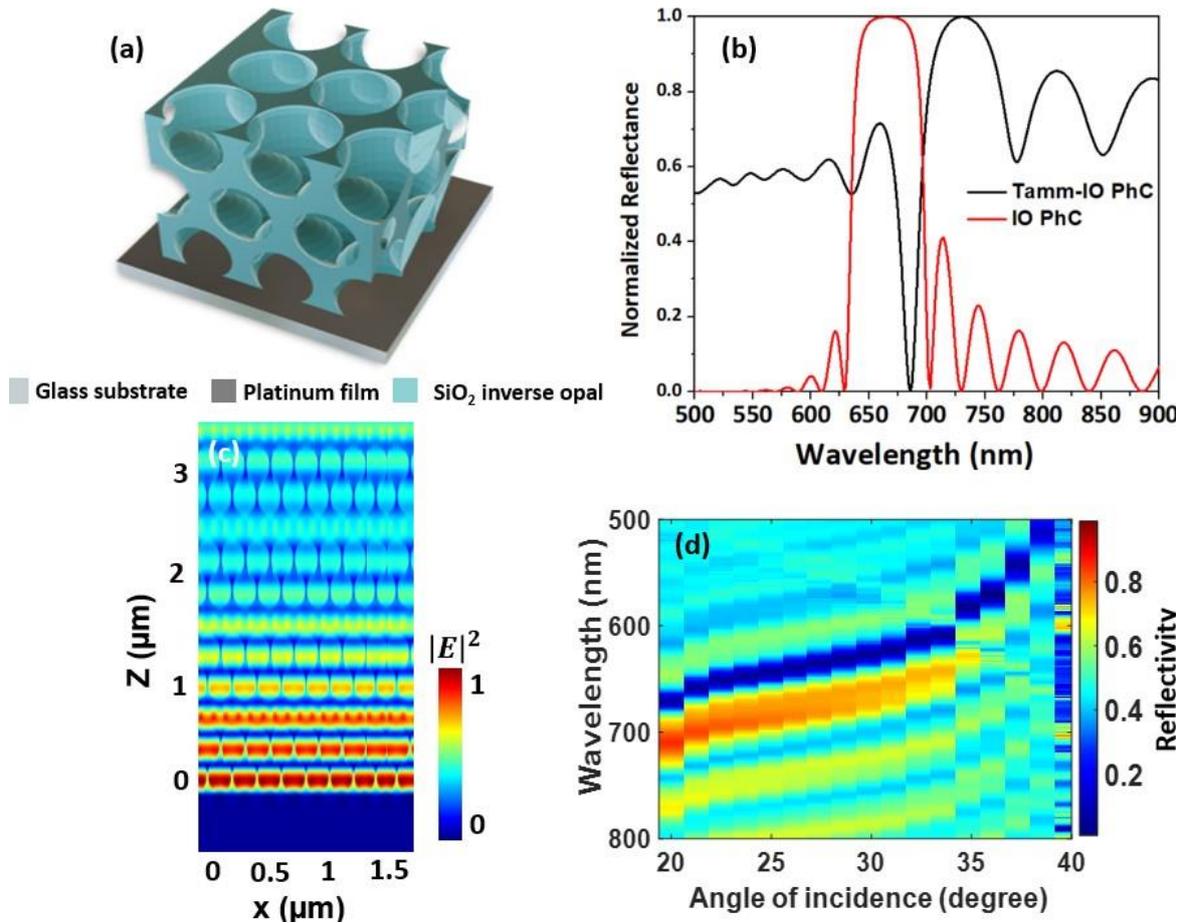

Fig.1: (a) Schematic of SiO$_2$ inverse opal on flat platinum coated vitreous silica substrate. (b) Simulated reflection spectra at normal incidence from inverse opal side with and without flat platinum layer. (c) Optical Tamm state electric field intensity distribution in Tamm-IO. (d) Computationally generated variable angle reflection spectra.

The design of the IO structure allows the sensing of various volatile organic compounds by infiltrating them in the void region of the structure. To computationally verify the dependence of the refractive index of the void region of the IO structure on its reflection spectrum, 3D-FDTD studies were carried out by varying the void refractive index from 1.00 to 1.37, with the refractive index of silica kept a constant. When the refractive index of the void was increased, the dip in the reflection spectra exhibited a redshift as shown in Figure 2(a). The electric field intensity at the resonance dip was computed to understand the origin of the dip. When the refractive index of the void was 1.00, the electric field intensity was found to be a maximum at the interface and decreased within the PhC structure. However, with the decrease in the refractive index contrast, the decay of the electric field is found to be less pronounced. The electric field distribution remains nearly unchanged, as observed in Figure 2(b), for the void refractive index increased to 1.37. The sensitivity of the Tamm-IO structure is calculated from the slope of the dotted line in Figure 2(a) to be ~ 254 nm/RIU. This sensitivity, denoted by $S$, is calculated using the formula,

$$S = \frac{\Delta \lambda}{\Delta n} \quad (2)$$

where, $\Delta \lambda$ represents the change in resonance wavelength and $\Delta n$ represents the change in refractive index of the void region.

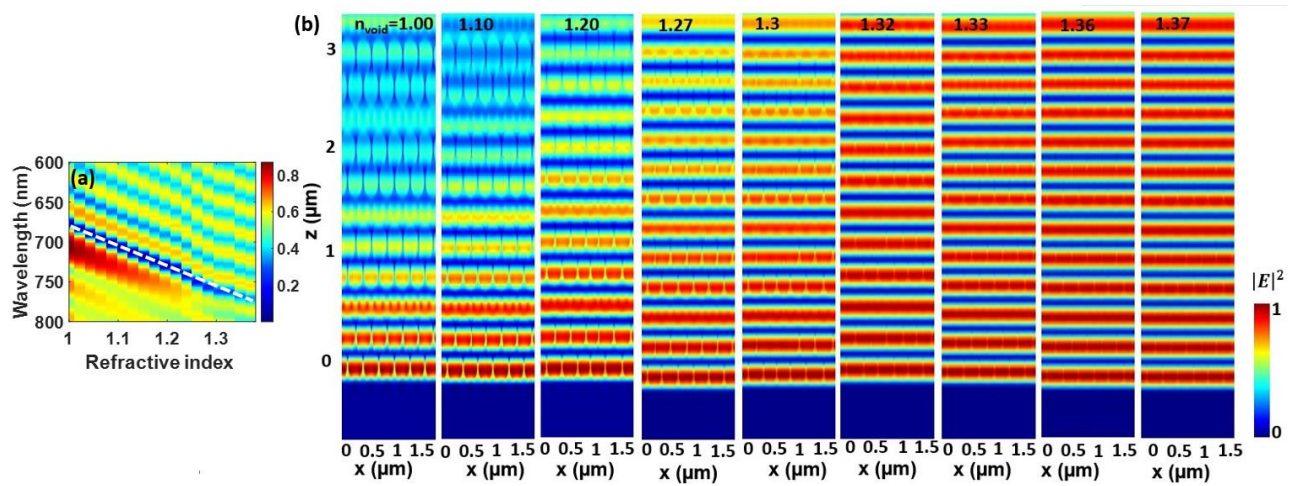

Fig. 2: (a) Computationally obtained normal incidence reflection spectra of Tamm-IO structure for different refractive indices of the void, $n_{void}$ = 1.00 to 1.37. (b) Electric field intensity distribution for six-unit cells of Tamm-IO structure for different refractive index of the void.

### 3. Device fabrication and characterization:

The schematic of the Tamm-IO PhC shown in Fig. 1(a) consists of $SiO_2$ inverse opal on Pt coated vitreous silica substrate. Silver or gold metallic thin films were not used due to crack formation or island formation during inverse opal fabrication. Platinum on the other hand is highly resistant to corrosion and oxidation, even at high temperatures, surpassing both gold and silver. In order to improve the adhesivity, a very thin film of thickness 5 nm of chromium was deposited before Pt on the vitreous silica substrate. To obtain the Tamm-IO PhC structure, the inverse silica opals were obtained by a modified co-assembly method [34], which involved the deposition of polymeric colloidal nanospheres from a suspension also containing a silica sol, with the aim of producing a silica matrix in the interstitial space of the polystyrene (PS) spheres that would be removed by calcination in a second step. In the specific a prehydrolized $SiO_2$ sol was obtained by mixing TEOS, the silica precursor, with water and ethanol. Briefly, 5 mL of water were mixed with 5 mL of ethanol and added to 1.25 mL of TEOS. The solution was maintained under vigorous stirring for 1 h. Then a uniform distribution of PS nanospheres in water, containing 150 μL of NPs, 5 mL of distilled water, and 80 μL of the above mentioned TEOS prehydrolized solution, was used to produce the colloidal film on the substrate. The metallic-dielectric substrate was vertically suspended in a vial containing the colloidal suspension. The slow evaporation of the solvents at 45 °C over a period of two days allowed the self-assembly of the colloidal particles at the meniscus and the consequent deposition of the colloidal crystal film on the substrate. Finally, an inverse silica opal was obtained by calcination of the PS template. The film was firstly heated for 2 h at 200 °C with a heating rate of 0.5 °C/min, and then for 2 h at 450 °C with a rate of 2 °C/min

For characterizing the device optically, reflection spectra acquired at normal incidence were taken using a fiber-optic UV–VIS spectrometer (Ocean Optics, USB4000, Largo, FL, USA), with a beam spot of about 1 $mm^2$ in area. Throughout the measurements, the fiber tip remained closely positioned to the sample surfaces. To standardize the data, all recorded spectra were normalized against a reference reflection spectrum obtained from an aluminium mirror, while variable reflectance measurements were obtained using Varian 5000 spectrophotometer. Structural characterization was carried out by means of scanning electron microscope (SEM) acquired using a Helios 5 PFIB CXe (ThermoFisher Scientific, Waltham, USA), to determine the range of the ordered domains and the dimension of the voids of the inverse opal.

Fig. 3(a) depicts the SEM image of the fabricated, highly ordered Tamm-IO PhC structure with the pore diameter of 365 ± 20 nm arranged in hexagonal lattice. The surface appears smooth and densely packed morphology, exhibiting a periodic arrangement of spherical void. The wall between the void appears smooth and interconnected, indicating well defined porous network. The experimentally obtained normal incidence reflection spectra of the IO and Tamm-IO structures, recorded from the IO side are shown in Fig 3(b). For the IO structure, the reflection peak occurs at 685 nm with FWHM of 48 nm. The reflection spectrum of the Tamm-IO exhibits a dip at $\lambda_{dip} \approx 687$ nm confirming the excitation of OTS. Coupling between the plasmonic mode of the flat metal and the Bloch mode of the IO photonic crystal gives rise to Tamm state when light is incident from the IO side [35,36]. Interestingly, the thickness of the metal layer does not influence the reflection spectra when measured from the IO side. The broadening of the resonance dips is attributed to the size distribution of PS particles used in the synthesis of IO structures, and also due to the defects arising during the self-assembly process.

Fig. 3(c) presents the experimentally measured reflection spectra at different angles of incidence for the Tamm-IO. In this study, the incident and reflection angles were systematically scanned from 20º to 40º relative to the normal to the sample surface. Notably, the OTS of the device exhibits a blue-shift as the angle of incidence increases, which can be attributed to the blue-shift in the bandgap of the inverse opal. A good agreement between the computational and experimental reflection spectra results at various angles of incidence confirming the high quality of the fabricated sample over a large area.

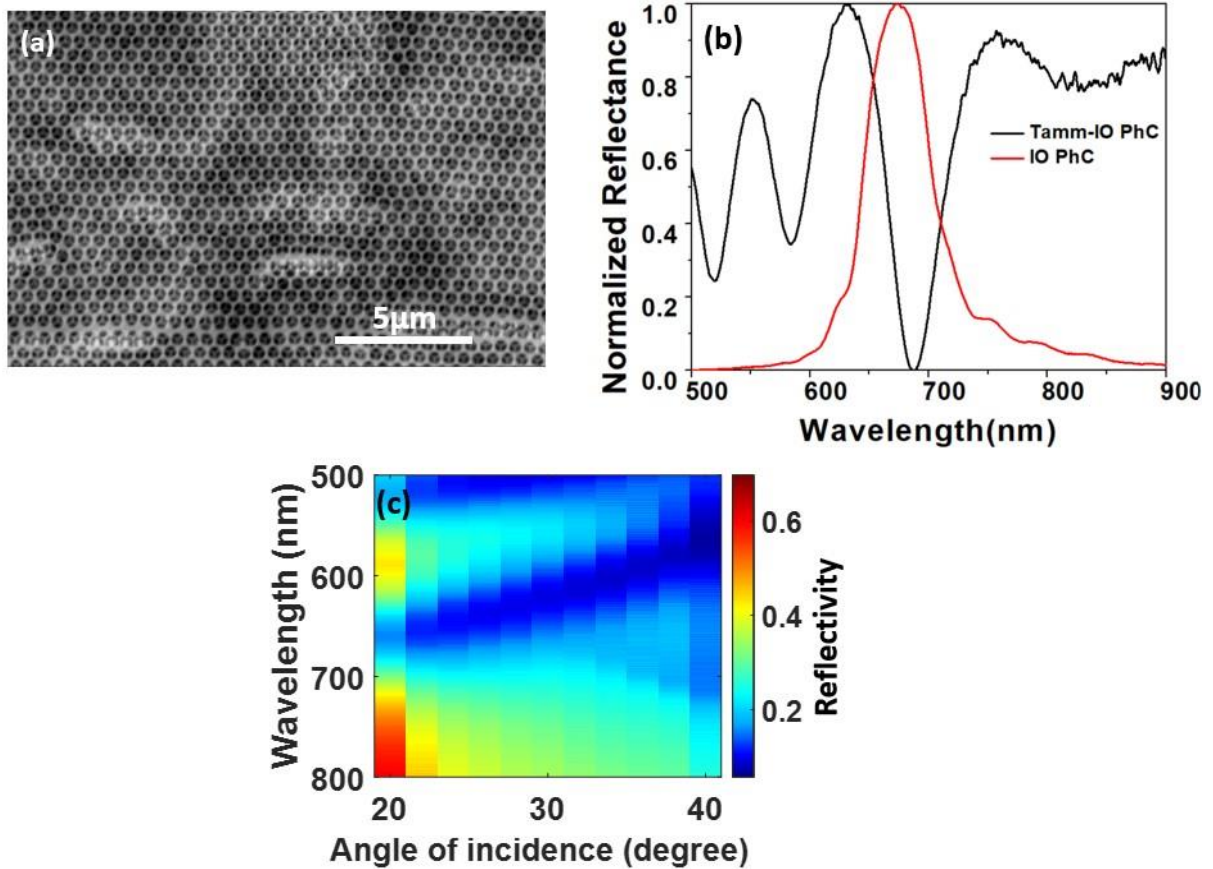

Fig. 3: (a) SEM image of SiO$_2$ inverse opal on Pt coated vitreous silica substrate. (b) Experimentally obtained normal incidence reflection spectra from Tamm-IO structure (black curve) and IO structure (red curve). (c) Measured reflection spectra from Tamm-IO at different angles of incidence.

### 4. Chemical sensing:

Volatile solvents that do not react with SiO$_2$ IO and platinum thin film were infiltrated into the Tamm-IO structure and the shift of the OTS was continuously monitored. For each solvent three separate trials were performed, by dispensing 1 µl of volatile organic solvent (ethanol and methanol) on the surface of the inverse opal structure. Fig. 4(a) shows the time-dependent normal incidence reflection spectra of Tamm-IO PhC structure before solvent infiltration, during infiltration and evaporation of the solvent. The reflection spectra were recorded for several seconds prior to the dispensing of the solvent at time $t_o$, to confirm the stability of the OTS. During the solvent infiltration and initial evaporation phase, the OTS was found to disappear and no signature of the OTS was observed for the duration $t$ when the solvent covered the Tamm-IO structure completely, as shown in fig. 4(a). Later, as the excesses solvent on the surface evaporated, the OTS reappeared with a spectral shift and gradually reverted back to its original spectral position, evident within the white dashed rectangle. Figs. 4(b-d) display the temporal dynamics of the OTS when ethanol was dispersed over the sample for three trials. Clearly, during ethanol evaporation, after time period $t$, the OTS reappears at a red-shifted position and later the OTS returns to its initial position after complete evaporation of ethanol. The duration for the shifted OTS to return to its original position is denoted as recovery time ($T$). Figs. 4(e-g) display the temporal response of the OTS when methanol was dispersed over the sample for three trials. The response of the OTS is similar to that observed for the case of ethanol infiltration. However, the recovery time of OTS is higher for methanol infiltration compared to ethanol, as tabulated in Table 1.

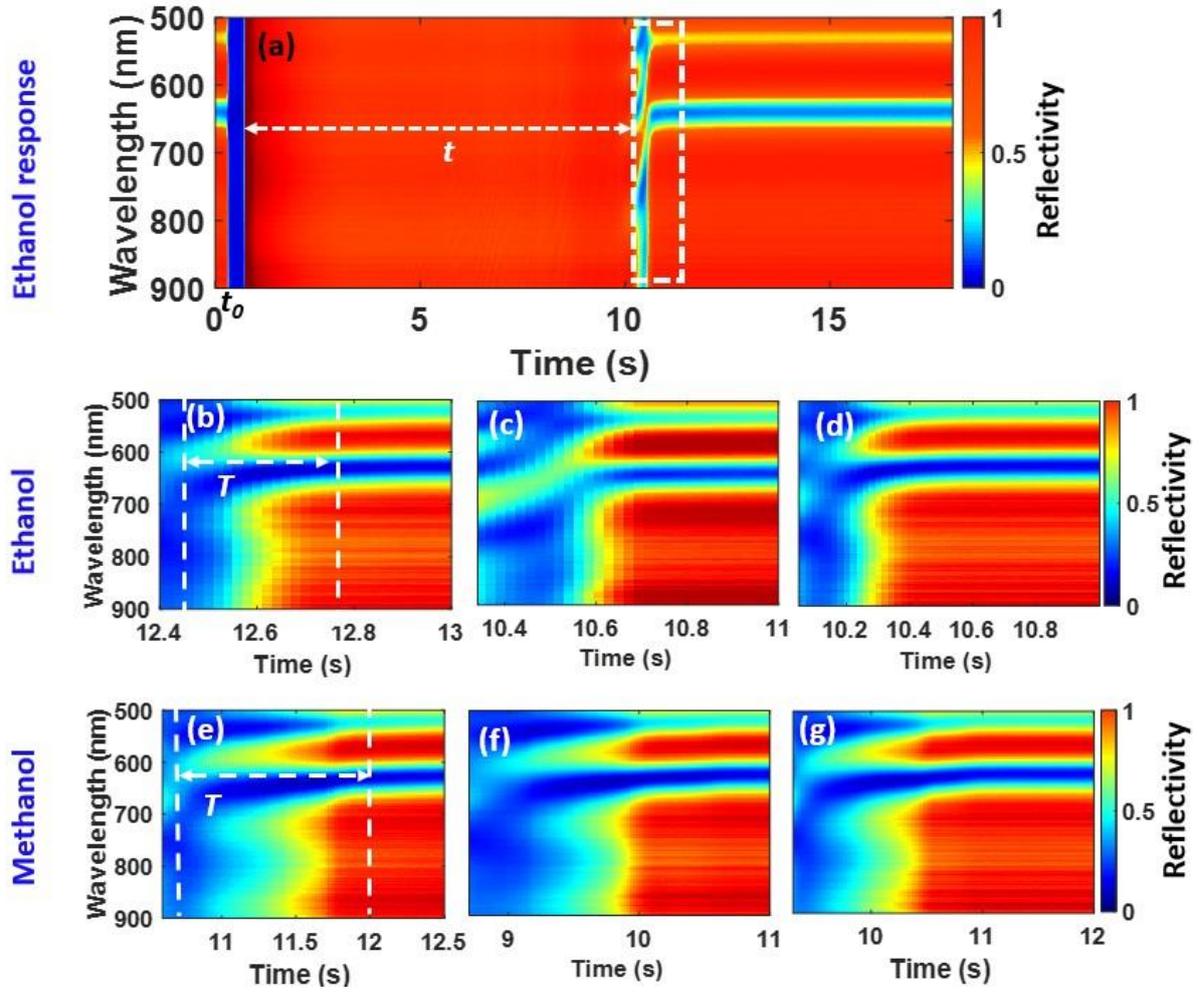

Fig. 4: Time dependent normal incidence reflection spectra of Tamm-IO PhC: (a) before solvent infiltration, during infiltration and evaporation of the solvent. (b-d) Display the temporal dynamics of the OTS when ethanol was dispensed over the sample for three trials. 4(e-g) The temporal response of the OTS when methanol was dispersed over the sample for three trials are presented. For all the six cases the recovery time of the OTS is denoted by $T$.

The duration for solvent evaporation is influenced by factors like vapor pressure, viscosity, diffusion rate, and surface tension of the solvent. The recovery time $T$ of the Tamm-IO for the case of ethanol, for three consecutive trials, is short compared to that for methanol, as observed in Table 1. This variation can be attributed to the surface tension and viscosity. The surface tension of ethanol and methanol are $\sim 21.97\ mN/m$ and $\sim 22.07\ mN/m$ [11], respectively. The lower surface tension of ethanol molecule allows it to overcome intermolecular forces more easily and escape into the air resulting in faster evaporation compared to methanol. Regarding viscosity, ethanol and methanol have values 1.107 mPa-s and 0.544 mPa-s, respectively [11]. The higher viscosity of ethanol prevents it from penetrating deeper into the Tamm-IO structure leading to its earlier recovery. In contrast, methanol with its lower viscosity, penetrates deeper into the porous structure and takes longer time to evaporate completely from the Tamm-IO PhC structure. The recovery times observed across three separate trials, are nearly identical, confirming the reliability of the device for repeatable measurements. The lower standard deviation errors in recovery times obtained for both the solvents confirm the repeatability of measurements.

Table 1: Recovery time of the Tamm-IO when infiltrated with volatile solvents, for three different trials.

| Solvents | Recovery time, $T$ (s) | | | Average $T$ (s) |
|---|---|---|---|---|
| | Trial 1 | Trial 2 | Trial 3 | |
| Ethanol | 0.3 | 0.4 | 0.5 | $0.4 \pm 0.1$ |
| Methanol | 2.0 | 1.9 | 2.0 | $2.0 \pm 0.6$ |

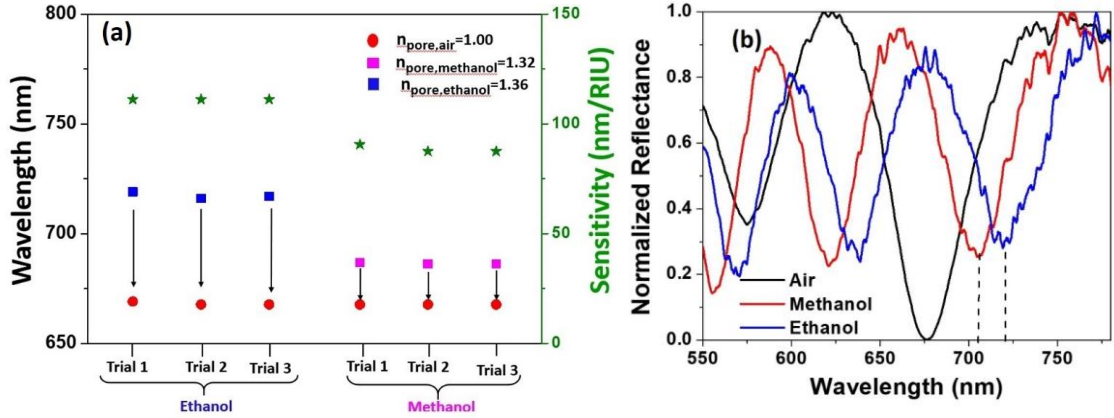

Fig.5: (a) Experimentally obtained Tamm resonance dip position when infiltrated with solvents (methanol and ethanol) and when solvents have fully evaporated, are presented. The sensitivity is also calculated for each trial for both the solvents. (b) Experimentally obtained reflection spectra from Tamm-IO structure when filled with air (black curve), methanol (red curve), ethanol (blue curve). The legends represent the medium in the void of Tamm-IO.

Figure 5(a) presents the experimentally observed spectral positions of the OTS at the boundaries of the interval $t$ for three different trials, wherein a red-shifted OTS is observed at the end of the time period $t$. The OTS shifts to higher wavelengths due to the changes in the void refractive index upon solvent infiltration, in agreement with the computational data presented in Fig. 2(a). After the solvent completely evaporates, the OTS returns to its initial position. The sensitivity of the device for each solvent trial is calculated using Eqn. 2 and presented in Fig. 5(a). Experimentally the sensitivity of the device is obtained as 110 nm/RIU, which is of the same order as that obtained computationally from Fig. 2(a). Experimentally, the viscosity of solvents hinders their deeper penetration into the Tamm-IO structure. The solvents with higher viscosity result in an earlier recovery of the OTS and prevent a further redshift of the OTS. This leads to lower sensitivity in experiments. In contrast, in the simulation, the refractive index of the void is completely changed to 1.36 and 1.32 for ethanol and methanol, respectively, which causes a significant shift of the OTS to higher wavelength, thereby increasing sensitivity. Figure 5(b) shows the experimentally obtained normalized reflection spectra of the Tamm-IO structure when the void is filled with air, methanol and ethanol. The spectra are plotted at $t_o$ for air-filled Tamm-IO, and at the end of the time interval $t$ for the two solvents. The shift of the OTS when the void is infiltrated with methanol and ethanol is evident, and the spectral shift is observed to be larger for the solvent with higher refractive index.

5. **Conclusion:**

In summary, our study focuses on the presence of OTS in the Tamm-IO structure, comprising $SiO_2$ IO on a flat metal substrate. Unlike traditional methods [28,37] involving corrosive chemicals for etching to create porous structures, we employed a straightforward fabrication approach. This method enhances the porosity of the three-dimensional photonic crystal, thereby increasing the surface area available for solvent or analyte penetration within the structure and facilitating access to the detection volume. The numerical FDTD simulations confirm the excitation of OTS at any angle of incidence as the dispersion relation of the mode lies entirely within the light cone. With an increase in the refractive index of the void the OTS exhibited a redshift, which has been verified by experiments. By infiltrating the Tamm-IO structure with methanol and ethanol solvents experimentally, we observed the redshift of OTS in the reflection spectra. The temporal dynamics of the OTS, its recovery time for both the solvents have been studied and its repeatability has been confirmed. The experimentally obtained sensitivity is found to be ~ 110 nm/RIU, which is in the same order as that predicted computationally. The sensitivity and repeatability of the Tamm-IO make them a potential platform for the development of integrated optic volatile organic compound sensors.

**Author contributions:**

**Category 1**

**Conception and design of study**: Rina Mudi, Andrea Chiappini, Kapil Debnath, Shivakiran Bhaktha B.N.

**Acquisition of data:** Rina Mudi, Alessandro Carpentiero, Monica Bollani, Mario Barozzi, Andrea Chiappini, Debnath, Shivakiran Bhaktha B N.

**Analysis and/or interpretation of data:** Rina Mudi, Andrea Chiappini, Kapil Debnath, Shivakiran Bhaktha B.N.

**Category 2**

**Drafting the manuscript:** Rina Mudi, Andrea Chiappini, Kapil Debnath, Shivakiran Bhaktha B.N.

**Revising the manuscript critically for important intellectual content:** Rina Mudi, Alessandro Carpentiero, Monica Bollani, Mario Barozzi, Kapil Debnath, Andrea Chiappini, Shivakiran Bhaktha B.N.

**Category 3**

**Approval of the version of the manuscript to be published (the names of all authors must be listed):** Rina Mudi, Alessandro Carpentiero, Monica Bollani, Mario Barozzi, Kapil Debnath, Andrea Chiappini, Shivakiran Bhaktha B.N.

**Declaration of competing interest**

The authors declare that they have no known competing financial interests or personal relationships that could have appeared to influence the work reported in this paper could have appeared to influence the work reported in this paper.

**Acknowledgments**

This research was carried out in the partly in the framework of Progetti@CNR EPOCALE funded by Consiglio Nazionale delle Ricerche. The authors acknowledge support from Science and Engineering Research Board via sponsored projects CRG/2020/002650, ECR/2018/000631, and Indian Space Research Organization project SAC/SEDA/EOSDIG/SSD/2023/1. The support received from MeitY, Indian Nanoelectronics Users Program (INUP) is also acknowledged.